%% file: main.tex
\documentclass{article}
\usepackage{spconf,amsmath,graphicx}
\usepackage{subcaption}
\usepackage[utf8]{inputenc} % allow utf-8 input
\usepackage[T1]{fontenc}    % use 8-bit T1 fonts
\usepackage{hyperref}       % hyperlinks
\usepackage{url}            % simple URL typesetting
\usepackage{booktabs}       % professional-quality tables
\usepackage{amsfonts}       % blackboard math symbols
\usepackage{nicefrac}       % compact symbols for 1/2, etc.
\usepackage{microtype}      % microtypography
\usepackage{xcolor}         % colors
\usepackage{multirow}

\def\ours{{DailyTalk}}

\title{DailyTalk: Spoken Dialogue Dataset for Conversational Text-to-Speech}
\name{Keon Lee$^{1,2}$\sthanks{Equal Contribution}\sthanks{This work was done when Keon Lee was with KAIST as a MS student.} \quad Kyumin Park$^{1*}$ \quad Daeyoung Kim$^{1}$} 
\address{$^{1}$School of Computing, KAIST, Rep. of Korea \\
$^{2}$KRAFTON Inc., Rep. of Korea \\
\texttt{keonlee@krafton.com, \{pkm9403, kimd\}@kaist.ac.kr}
}
\begin{document}
%\ninept
%
\maketitle
\input{content/00_abstract}
\input{content/01_introduction}
\input{content/02_data_construction}
\input{content/04_baseline}
\input{content/05_experiments}
\input{content/07_conclusion}
\input{content/99_acknowledge}

% References should be produced using the bibtex program from suitable
% BiBTeX files (here: strings, refs, manuals). The IEEEbib.bst bibliography
% style file from IEEE produces unsorted bibliography list.
% -------------------------------------------------------------------------
\bibliographystyle{IEEEbib}
\bibliography{bibtex}

\end{document}

%% file: content/00_abstract.tex
\begin{abstract}
The majority of current Text-to-Speech (TTS) datasets, which are collections of individual utterances, contain few conversational aspects. In this paper, we introduce \ours{}, a high-quality conversational speech dataset designed for conversational TTS. We sampled, modified, and recorded 2,541 dialogues from the open-domain dialogue dataset DailyDialog inheriting its annotated attributes. On top of our dataset, we extend prior work as our baseline, where a non-autoregressive TTS is conditioned on historical information in a dialogue. From the baseline experiment with both general and our novel metrics, we show that \ours{} can be used as a general TTS dataset, and more than that, our baseline can represent contextual information from \ours{}. The \ours{} dataset and baseline code are freely available for academic use with CC-BY-SA 4.0 license \footnote{\url{https://github.com/keonlee9420/DailyTalk}}.
\end{abstract}


%% file: content/01_introduction.tex
\section{Introduction}
\label{sec:intro}

% 대화할 때 맥락을 파악하는 점이 중요하지만, 현재 이를 만들 데이터셋이 없음
Being a part of the conversation process, a TTS system needs to maintain and express the context related to the current dialogue. However, previous TTS models have limitations in context representation since they perceive each utterance separately regardless of the dialogue \cite{ito2017ljspeech, zen2019libritts}. Some recent studies proposed context-aware TTS models \cite{guo2021conversational, yan2021adaspeech3, xue2022paratts}, but they trained the model with an internal dataset, which is not open to the public. Also, some corpora collected from real-world conversations or acts are publicly available as \cite{poria2019meld}, but they have several limitations, including background noise or inconsistent record quality.

% DailyDialog를 녹음한 DailyTalk을 만듦
To resolve this, we introduce \ours{}, a high-quality dialogue speech dataset for TTS. Deriving dialogues from DailyDialog dataset \cite{li2017dailydialog}, we construct a new speech dataset by processing and recording selected dialogues. We design \ours{} to ensure both general and conversational speech synthesis quality: studio-quality audio, simultaneous recording of two participants, and adding filling-gap words (\textit{uh}, \textit{umm}) in the part of the dataset. At the same time, we preserve valuable characteristics of the DailyDialog dataset: academically open license and various dialogue annotations.

Together, we show a baseline model for conversational TTS to investigate the effect of contextual information when synthesizing the speech. Based on FastSpeech2 \cite{ren2021fastspeech2} architecture, the baseline consumes dialogue history as an additional input  following \cite{guo2021conversational}. Using our baseline, we show training model with our dataset can synthesize natural speech in the scope of both single utterances and whole dialogue.

Our major contributions are as below.  
\begin{enumerate}
    \item We provide the first open dataset for conversational TTS with a working baseline. 
    \item We suggest the entire pipeline to build a conversational TTS system from the text dialogue dataset. 
    \item We suggest several evaluation criteria to evaluate conversational TTS. 
\end{enumerate}
% 어떻게 만들고, 어떻게 증명하는지
In the following sections, we introduce our conversational TTS dataset with recording details, baseline, and experiments. 

%% file: content/02_data_construction.tex
\section{DailyTalk}

Construction pipeline of \ours{} consists of pre-record processing, recording, and post-record processing while taking advantage of details suggested by  \cite{guo2021conversational}. To guarantee sufficient data size and preserve dialogue characteristics, we assume the conversation is dyadic between one male and one female speaker.
Details are introduced in the following sections.

\subsection{Pre-record Processing}
Pre-record processing was the step for selecting and refining dialogues of \cite{li2017dailydialog} in two criteria: turn length and participant information. The selected dialogues have more than 5 turns because short dialogues are less likely to have context. Next, we only changed the characters' names to minimize the gap from the original dataset. In some dialogues with two participants of the same gender, we attempted to change one participant to another gender unless the other annotations (emotion, speech act, topic) were maintained. We excluded the dialogue from our dataset when the gender change harms the other annotations.

\subsection{Recording}
In the recording step, we managed conditions to provide a clear, conversational audio dataset. Two English-fluent speakers were employed as voice actors, each having lived in the US for at least 3 years. The actors recorded actual conversations rather than just reading the script, and they were directed to follow each utterance's emotion label. Additionally, we ordered them to add filling gaps (i.e. \textit{uh}, \textit{um}) in approximately half of the dialogues. This direction enables the TTS model to express filling gaps as well. The entire recording is done in a studio.

\subsection{Post-record Processing}
Post-record processing is conducted to correct misalignment between speeches and scripts. Since we directed voice actors not to act exactly to script, recorded speech might differ from the script in aspects of filling gaps or some conversational words. For this task, 6 annotators fluent in English are hired to fix incorrect transcription for this task.

\subsection{Statistics}
Table~\ref{tab:statistics_tts} shows detailed statistics of \ours{}, containing 2,541 dialogues which is 20 hours in total. We kept all script-related label distribution (e.g., emotion, dialog act, topic) derived from DailyDialog.\footnote{Detailed statistics are provided in \url{https://github.com/keonlee9420/DailyTalk}}

\input{supplementary/statistics_count}

%% file: supplementary/statistics_count.tex
% \begin{table*}[]
%     \centering
%     \begin{tabular}{l|cc|ccc}
%     \toprule
%         \multirow{2}{*}{\textbf{Dataset}} & \multirow{2}{*}{\textbf{LJSpeech}} & \multirow{2}{*}{\textbf{VCTK}} & \multicolumn{3}{c}{\textbf{\ours{}}} \\
%          & & & Male & Female & Total \\
%     \midrule
%         \# clips & 13,100 & 44,283 & 11,867 & 11,906 & 23,773 \\
%         \# words & 249,637 & 387,030 & 124,624 & 126,721 & 251,345 \\
%         \# characters & 1,310,342 & 1,728,414 & 516,055 & 524,744 & 1,040,799 \\
%         total duration (s) & 86117 & 149252 & 38902 & 39124 & 78026 \\
%         mean duration per clip (s) & 6.574 & 3.370 & 3.278 & 3.286 & 3.282 \\
%         mean \# phoneme per clip & 70.817 & 27.292 & 29.471 & 29.884 & 29.678 \\
%         \# distinct words & 14,946 & 5,775 & 7,220 & 7,329 & 10,160 \\
%         % \# dialogues & - & - & 2,541 & 2,541 & 2,541 \\
%         % mean turns per dialogue & - & - & 4.670 & 4.686 & 9.356 \\ 
%         mean $f_0$ & 127.490 & 63.384 & 96.199 & 142.133  & 119.258 \\
%         std $f_0$ & 108.778 & 88.021 & 81.047 & 112.733 & 100.889 \\
%     \bottomrule
%     \end{tabular}
%     \caption{Comparison of \ours{} to existing TTS datasets. Statistics of VCTK is an aggregation of entire 110 speakers, where LJSpeech consists of single-speaker data.}
%     \label{tab:statistics_tts}
% \end{table*} 

\begin{table}[t]
    \centering
    \begin{tabular}{l|ccc}
    \toprule
        \textbf{feature} & \textbf{Male} & \textbf{Female} & \textbf{Total} \\
    % \multicolumn{3}{c}{} \\ Male & Female & Total \\
    \midrule
        \# clips & 11,867 & 11,906 & 23,773 \\
        \# words & 124,624 & 126,721 & 251,345 \\
        % \# characters & 516,055 & 524,744 & 1,040,799 \\
        total duration (s) & 38902 & 39124 & 78026 \\
        mean duration/clip (s) & 3.278 & 3.286 & 3.282 \\
        mean \# phone/clip & 29.471 & 29.884 & 29.678 \\
        \# distinct words & 7,220 & 7,329 & 10,160 \\
        \# dialogues & 2,541 & 2,541 & 2,541 \\
        mean turns/dialogue & 4.670 & 4.686 & 9.356 \\
        \# dialogues w/ fgs & 962 & 689 & 1,452 \\
        % mean $f_0$ & 96.199 & 142.133  & 119.258 \\
        % std $f_0$ & 81.047 & 112.733 & 100.889 \\
    \bottomrule
    \end{tabular}
    \caption{Detailed stats of \ours{}. Filling gaps (fgs) indicates  \textit{uh} and \textit{umm}.}
    \label{tab:statistics_tts}
\end{table} 

%% file: content/04_baseline.tex
\section{Baseline}
\label{sec:baseline}
\input{supplementary/model_architecture}

As shown in Figure~\ref{fig:baseline_architecture}, our baseline model for conversational TTS consists of a context encoder added to improved FastSpeech2 \cite{ren2021fastspeech2}. The FastSpeech2 backbone is augmented by replacing Montreal Forced Aligner \cite{mcauliffe2019mfa} with a CTC-based aligner \cite{badlani2021one} so that duration is learned during training in an end-to-end manner. The conversational context encoder is from \cite{guo2021conversational} to condition on historical information. The reason for dropping Tacotron2 \cite{guo2021conversational} as a backbone is difficulty aligning when training data becomes expressive \cite{lee21h_interspeech}. Our implementation is based on two open-sourced projects where the Transformer \cite{vaswani2017attention} encoder and decoder are connected by the unsupervised duration modeling \footnote{https://github.com/keonlee9420/Comprehensive-Transformer-TTS}, and where the conversational context encoder from \cite{guo2021conversational} is applied to FastSpeech2 model \footnote{https://github.com/keonlee9420/Expressive-FastSpeech2}.

%% file: supplementary/model_architecture.tex
% \begin{figure}
%     \centering
%     \includegraphics[width=\textwidth]{images/model_temp.png}
%     \caption{Baseline model architecture. Left image is overall architecture of our baseline. Center image depicts details of unsupervised alignment module in the baseline. Right is the detailed architecture of context encoder in the baseline.}
%     \label{fig:baseline_architecture}
% \end{figure}

\begin{figure}[t]
    \centering
    \includegraphics[width=0.7\columnwidth]{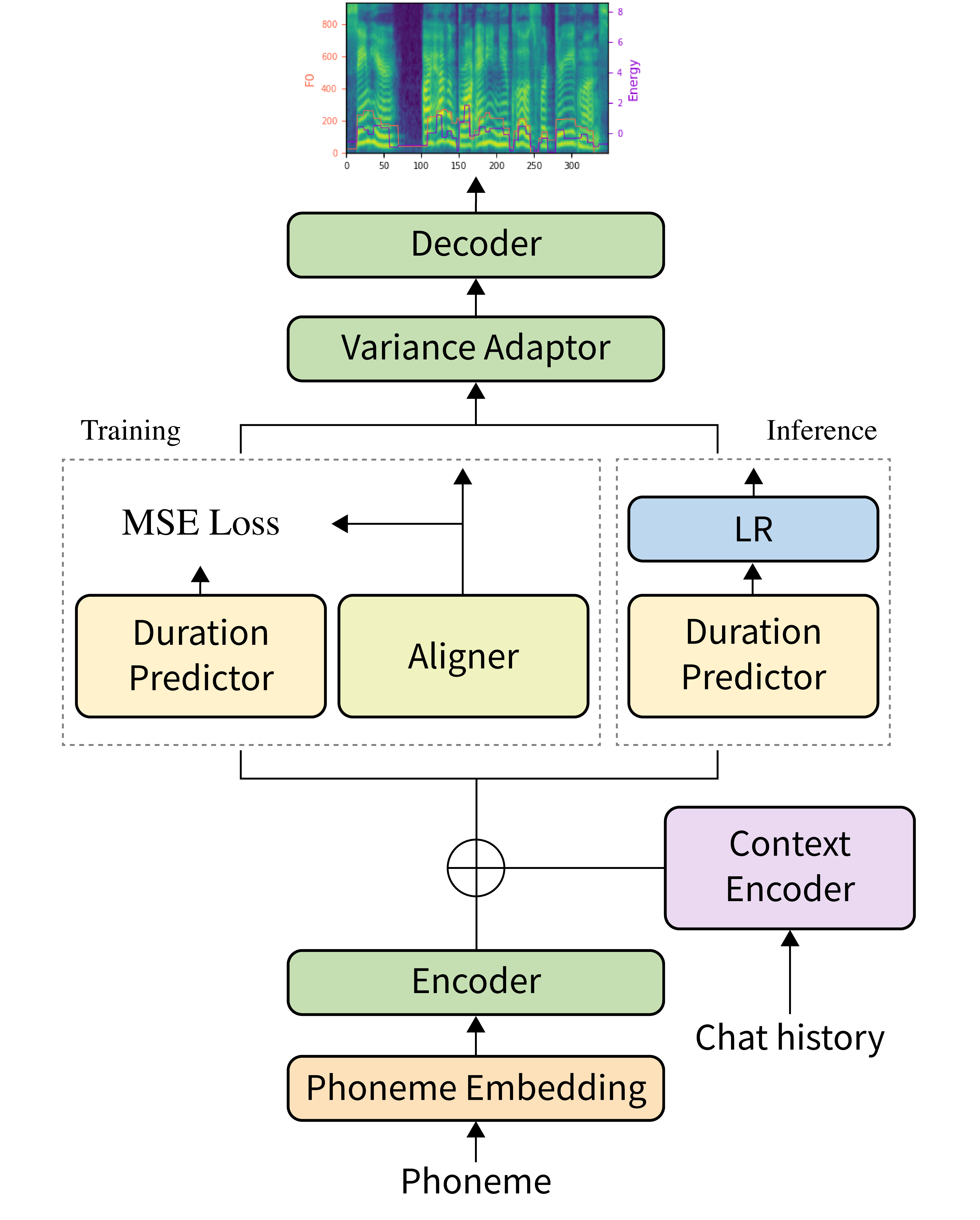}
    \caption{Baseline model architecture.}
    \label{fig:baseline_architecture}
\end{figure}

%% file: content/05_experiments.tex
\section{Experiments}
\label{sec:experiments}

We conduct two experiments to verify the efficacy of \ours{} and provide the baseline. The first experiment compares our dataset with existing datasets, which proves the integrity of ours for neural TTS. Next, we show the performance of the context-aware TTS model trained with \ours{} to prove the effectiveness in the aspect of context modeling. The models used in experiments are trained with 16 batch sizes using a single NVIDIA RTX 3090. Training hyperparameters such as optimizers are based on each model's original configuration. We stopped the training when the validation loss was no longer dropped. When we trained a model on LJSpeech, we randomly selected 512 sentences as a validation set. For VCTK, we selected two sentences per speaker for validation. In \ours{}, we set 128 dialogues for the validation set. We randomly sampled 260 utterances (of 15 dialogues) for all the following experiments. We measure all scores with a confidence interval of 95\%.

\subsection{Integrity as a TTS dataset}
\label{sec:experiment1}
Since the primary goal of \ours{} is training TTS models, TTS models should be successfully trained on our dataset. To show this, we train three neural TTS models: Tacotron2 \cite{shen2018natural},  FastSpeech2 \cite{ren2021fastspeech2} and our baseline. Tacotron2 and FastSpeech2 are selected as representative models of autoregressive TTS and non-autoregressive TTS. Note that our baseline can be trained on \ours{} only since it requires chat history as an input. We randomly split the training set and test set and used only the training set for model training, where test sets are used for further experiments.

To measure the performance of each model trained on each dataset, we conducted a Mean Opinion Score (MOS) test for synthesized and reconstructed speeches. We asked people how natural each synthesized speech is by selecting in a range from 1 (Completely not Human-like) to 5 (Completely Human-like). Only people from the US participated in the test since English-fluent people are required for this evaluation. We used data in the test set to synthesize speech when conducting the MOS test. We asked 30 different people to measure the naturalness of each audio, then calculated the average score for each model-dataset pair. 

\input{supplementary/experiment1_result}

Table~\ref{tab:exp1_result} shows the MOS result of each model-dataset pair. Overall, synthesized speeches from models trained on \ours{} sound as natural as other datasets. In reconstructions, \ours{} scores between LJSpeech and VCTK, which is coherent considering the number of speakers. Scores of the synthesized speeches from models (Tacotron2, FastSpeech2) show within the confidence interval for all three datasets. Although our baseline reports reasonable scores, it is still lower than the others as a result of increased complexity \cite{lee21h_interspeech}. Note that the lower score does not imply the usefulness of the chat history, which we will explore on in the next section.

\subsection{Conversational TTS}
\label{sec:experiment2}
\input{supplementary/experiment2_result}
\input{supplementary/experiment2_divided}

The major difference between our dataset from the others is the conversational information. Aiming to show how the conversational TTS model differs from general TTS models, we design several evaluation tasks that ask naturalness of dialogue in terms of context preservation. Our assumption throughout the evaluations is that people can be aware of context through the relation between adjacent utterances. We design the following four novel metrics for our experiments: Varying utterance groups given to evaluators.

\begin{itemize} 
    \item Dialogue-level MOS: A set of audio is an entire dialogue, where the TTS model synthesizes all utterances in the dialogue.
    \item Pairwise MOS (GT-synthesized): A set of audio is a pair of adjacent utterances, where the prior utterance is ground truth, and the latter is synthesized from the model.
    \item Pairwise MOS (Synthesized pair): A set of audio is a pair of adjacent utterances, where both utterances are synthesized from the model. 
    % \item Utterance-level MOS: A set of audio is single utterance, which is synthesized from the model.
    \item User-System Simulation MOS: A set of audio is an entire dialogue, where utterances of one speaker are ground truth and of the other speaker are synthesized from the model.
\end{itemize}

In each evaluation, we asked people from the US how natural the spoken dialogue is in terms of context maintenance. We conduct both MOS and Comparative MOS (CMOS) throughout all evaluations. In the MOS test, people evaluate the naturalness in the 1-5 range. In the CMOS test, people listen to two sets of audio synthesized from different models with the same script, then select a preference score from -3 (A is much better) to 3 (B is much better). In all experiments, complete scripts of each dialogue are provided to participants to evoke the context of the given audio set during the test.

% 결과 정리
Table~\ref{tab:exp2_result} reports the experiment results. Even FastSpeech2 obtains higher naturalness score in section~\ref{sec:experiment1}, our baseline reports higher score than FastSpeech2 in three experiments: dialogue-level, User-System and Pairwise(synth pair) MOS tests. In other words, the audio quality is ranked a bit lower only when the output speech from two models is compared directly. But when it is evaluated under conversational conditions, our baseline beats the competitor for all metrics, showing the effectiveness of chat history, except the GT-synth pair. Notably, Pairwise (GT-synth) shows a different tendency compared to other metrics, so we discuss it in section~\ref{sec:turn_division}. 

Pairwise (GT-synth) and pairwise CMOS tests also show decreased gap. Dialogue-level and User-System CMOS results still show a gap that FastSpeech2 synthesizes better than ours. We suspect that listeners are still affected by audio quality even with the guideline. Considering dialogue-level and user-system CMOS shows a similar gap, the gap tends to increase when people listen to a larger amount of audio. This gap may also come from a failure in a single utterance. Even a little mistake in a single utterance does not affect overall context-awareness much, and it may influence a lot compared to non-failure results. Therefore, we concluded that CMOS results would be said to be affected more by speech quality than MOS results, not context-awareness.

Summing up, Dialogue-level criteria can evaluate whether the model synthesizes the entire dialogue correctly. The pairwise criterion would assess the relationship between adjacent utterances, such as context preservation. In pairwise evaluation, the pairwise (GT-synth) metric can be used when synthesized speech needs to resemble ground truth recording. The pairwise (synth pair) test can be used to evaluate whether two synthesized speeches contain a shared context. User-System evaluation can be used at the application level, where the system replies user's utterance, such as voice secretary.

\subsection{Turn Division}
\label{sec:turn_division}
Usually, contextual information piles up while the dialogue continues. Therefore, it is hard to represent the context in the beginning turns of the dialogue. To show context piling up and the apparent difference in context modeling, we divide the MOS result into half: before/after turn 5. We set the threshold to 5 since the average number of turns per dialogue is 9.3 (see Table~\ref{tab:statistics_tts}) and turn 5 is the nearest half-point. 

Table~\ref{tab:exp2_divided} shows the divided result of MOS test. In both experiments, the MOS result of our baseline remains consistent or increases after turn 5, while the score of FastSpeech2 decreases or increases but is lower than ours. Interestingly, following GT doesn't guarantee that the model will output more conversationally natural speech. And still, we can assess the performance of the conversational TTS model independently from the similarity to GT. This is proved by several findings from the results as follows. First, the CMOS in GT-synth pairwise reveals that the baseline goes far from the GT after 5 turns, more than FastSpeech2. The CMOS in the synth pair, on the other hand, shows better performance in ours after 5 turns. And this also explains the result of Pairwise(GT-synth) in Table~\ref{tab:exp2_result}. Second, although the performance of ours doesn’t increase after turn 5 in Pairwise (GT-synth) and that of FastSpeech2 increases after turn 5, we would like to highlight that the performance degradation in GT-synth after turn 5 is lower, but the gain in synth pair is huge in ours.

% \subsection{More features}
% We leave modeling filling gaps such as 'uh' and 'umm' for future work: expanding candidate features with a more precise definition of spontaneous conversation. Furthermore, expanding the dataset would be another future work, including a multispeaker version with more than two speakers or a multilingual version considering the cultural differences in conversation.

%% file: supplementary/experiment1_result.tex
% \begin{table*}[t]
%     \centering
%     \begin{tabular}{c|cccc}
%     \toprule
%         \multirow{2}{*}{Dataset} & \multicolumn{4}{c}{Models} \\
%          & GT-recon & Tacotron2 & FastSpeech2 & Baseline\\
%     \midrule
%         LJSpeech & 3.900 $\pm$ 0.067 & 3.712 $\pm$ 0.068 & 3.852 $\pm$ 0.070 & \\
%         VCTK & 3.807 $\pm$ 0.071 & 3.723 $\pm$ 0.075 & 3.753 $\pm$ 0.072 & \\
%         \ours{} & 3.827 $\pm$ 0.064 & 3.697 $\pm$ 0.070 & 3.818 $\pm$ 0.066 & 3.688 $\pm$ 0.077 \\
%     \bottomrule
%     \end{tabular}
%     \caption{TTS integrity test result. GT-recon refers to the speeches reconstructed from ground truth mel-spectrogram. Baseline is a model suggested in section~\ref{sec:baseline}. Tacotron2 and FastSpeech2 are neural TTS models for comparison.}
%     \label{tab:exp1_result}
% \end{table*}

\begin{table}[ht]
    \centering
    \begin{tabular}{c|ccc}
    \toprule
        \textbf{Model} & \textbf{LJSpeech} & \textbf{VCTK} & \textbf{\ours{}} \\
    \midrule
        GT-Recon & 3.90 $\pm$ 0.07 & 3.81 $\pm$ 0.07 & 3.83 $\pm$ 0.06 \\
        Tacotron2 & 3.71 $\pm$ 0.07 & 3.72 $\pm$ 0.08 & 3.70 $\pm$ 0.07 \\
        FastSpeech2 & 3.85 $\pm$ 0.07 & 3.75 $\pm$ 0.07 & 3.82 $\pm$ 0.07 \\
        Baseline & & & 3.69 $\pm$ 0.08 \\
    \bottomrule
    \end{tabular}
    \caption{TTS integrity test result. GT-recon refers to the speeches reconstructed from ground truth mel-spectrogram. Baseline is a model suggested in section~\ref{sec:baseline}.}
    \label{tab:exp1_result}
\end{table}

%% file: supplementary/experiment2_result.tex
% \begin{table*}[t]
%     \centering
%     \begin{tabular}{c|cccc}
%     \toprule
%         \multirow{2}{*}{Model} & \multicolumn{4}{|c}{Experiment} \\
%          & Dialogue-level & Pairwise(GT-synth) & Pairwise(synth pair) & User-System \\
%     \midrule
%         \multicolumn{5}{c}{MOS} \\
%     \midrule
%         FastSpeech2 & 3.765 $\pm$ 0.051 & \textbf{3.767} $\pm$ 0.035 & 3.864 $\pm$ 0.035 & 3.673 $\pm$ 0.068 \\
%         Ours & \textbf{3.787} $\pm$ 0.051 & 3.737 $\pm$ 0.035 & \textbf{3.898} $\pm$ 0.033 & \textbf{3.697} $\pm$ 0.071 \\
%     \midrule
%         \multicolumn{5}{c}{CMOS} \\
%     \midrule
%         Ours-FS2 & -0.294 $\pm$ 0.147 & -0.066 $\pm$ 0.070 & -0.066 $\pm$ 0.065 & -0.223 $\pm$ 0.122 \\
%     \bottomrule
%     \end{tabular}
%     \caption{Result of experiments in section~\ref{sec:experiment2}. Above two rows report MOS test result of each model with confidence interval, where Ours indicates our baseline model. The last row shows CMOS test result between our baseline and FastSpeech2 (FS2). Positive CMOS score means that our baseline scores better than FastSpeech2, and negative score is vice versa.}
%     \label{tab:exp2_result}
% \end{table*}

\begin{table}[t]
    \centering
    \begin{tabular}{c|cc||c}
    \toprule
        \multirow{2}{*}{\textbf{Experiment}} & \multicolumn{2}{|c}{\textbf{MOS}} & \textbf{CMOS} \\
         & \textbf{FS2} & \textbf{Ours} & \textbf{Ours-FS2} \\
    \midrule
        Dialogue & 3.77 $\pm$ 0.05 & \textbf{3.79} $\pm$ 0.05 & -0.29 $\pm$ 0.15 \\
        Pairwise & \multirow{2}{*}{\textbf{3.77} $\pm$ 0.04} & \multirow{2}{*}{3.74 $\pm$ 0.04} & \multirow{2}{*}{-0.07 $\pm$ 0.07} \\
        (GT-synth) & & & \\
        Pairwise & \multirow{2}{*}{3.86 $\pm$ 0.04} & \multirow{2}{*}{\textbf{3.90} $\pm$ 0.03} & \multirow{2}{*}{-0.07 $\pm$ 0.07} \\
        (synth pair) & & & \\
        User-Sys & 3.67 $\pm$ 0.07 & \textbf{3.70} $\pm$ 0.07 & -0.22 $\pm$ 0.12 \\
    \bottomrule
    \end{tabular}
    \caption{Result of experiments in section~\ref{sec:experiment2}.
    %Left two columns report MOS test result of each model with confidence interval, where Ours indicates our baseline model. The right column shows CMOS test result between our baseline and FastSpeech2 (FS2). 
    Positive CMOS score means that our baseline scores better than FastSpeech2, and negative score is vice versa.}
    \label{tab:exp2_result}
\end{table}

%% file: supplementary/experiment2_divided.tex
\begin{table*}[t]
    \centering
    \begin{tabular}{c|cc}
    \toprule
        \multirow{2}{*}{\textbf{Model}} & \multicolumn{2}{|c}{\textbf{Experiment}} \\
         & \textbf{Pairwise(GT-synth)} & \textbf{Pairwise(synth pair)} \\
    \midrule
        \multicolumn{3}{c}{\textbf{MOS}} \\
    \midrule
        FastSpeech2 & 3.817 $\pm$ 0.051 / 3.725 $\pm$ 0.049 & 3.842 $\pm$ 0.050 / 3.882 $\pm$ 0.047 \\
        Ours & 3.738 $\pm$ 0.051 / 3.737 $\pm$ 0.048 & 3.854 $\pm$ 0.048  / 3.937 $\pm$ 0.046 \\
    \midrule
        \multicolumn{3}{c}{\textbf{CMOS}} \\
    \midrule
        Ours-FastSpeech2 & 0.003 $\pm$ 0.102 / -0.124 $\pm$ 0.097 & -0.145 $\pm$ 0.095 / 0.001 $\pm$ 0.089  \\
    \bottomrule
    \end{tabular}
    \caption{Separated MOS scores of Pariwise(GT-synth), Pairwise(synth pair) experiments. 
    %In each cell, score in front is a MOS score before fifth turn, and the latter score is the average score of remaining turns. 
    Note that we cannot conduct turn division experiments for dialogue-level and User-System evaluations, because those methods evaluate conversation in dialogue-unit. Since they are not evaluated by each turn, evaluation cannot be divided into turns. }
    \label{tab:exp2_divided}
\end{table*}

%% file: content/07_conclusion.tex
\section{Conclusion}
\label{sec:conclusion}

In this paper, we introduce \ours{}, the first public dataset for conversational TTS for modeling various features and situations in conversation. Together, we present a baseline with pretrained models provided. Throughout the experiment, we prove that our dataset can be used to train several types of models, including general TTS and conversational TTS. At the same time, we propose several evaluation methods for the context representation of conversational TTS so that people can choose appropriate metrics for their needs.
Future works may focus on expanding dataset with more spontaneous features, speakers, and languages considering the cultural diversity.

%One may point out that only including filling words such as 'uh' and 'umm' is not enough to model spontaneous dialogue. This limitation of our dataset remains in future work: expanding candidate features in a spontaneous dataset with a more precise definition of spontaneous conversation. Furthermore, developing the dataset would be another future work, including a multispeaker version with more than two speakers or a multilingual version considering the cultural differences in conversation.

%% file: content/99_acknowledge.tex
\section{Acknowledgement}

This research was supported by the MSIT(Ministry of Science and ICT), Korea, under the Grand Information Technology Research Center support program(IITP-2022-2020-0-01489), Operation of SmartCity Lab based on Digital Twin (Grant 2022-0-00407) and Project-based AI Talent Fostering Program (RS-2022-00143911) supervised by the IITP(Institute for Information \& communications Technology Planning \& Evaluation)